\documentclass[]{spie}  

\usepackage{amsmath,amsfonts,amssymb}
\usepackage{graphicx}
\usepackage[colorlinks=true, allcolors=blue]{hyperref}

\title{First results of a continuous monitoring campaign of the PLATO Southern field}

\author[a]{Thomas Granzer}
\author[a]{Jorg Weingrill}
\author[a]{Klaus G. Strassmeier}
\author[a]{Arto Jarvinen}
\affil[a]{Leibniz Institut f{\"u}r Astrophysik Potsdam (AIP), Potsdam, Germany}

\authorinfo{Further author information: (Send correspondence to T.G.)\\T.G.: E-mail: tgranzer@aip.de}

\begin{document} 
\maketitle

\begin{abstract}
The BMK10k project started as an ancillary project to the PLATO ESA mission, dedicated to photometrically monitor the PLATO Southern field, to help with source confusion, and to mitigate the problem of false-positive exoplanet detections. Planned as a long-term project, the BMK10k should see an operational time-scale in the order of a decade, well above the proposed PLATO operational phase. Thus, BMK10k may help resolving single-event transit issues detected with PLATO and may be in the unique situation to transit-detect cold Jupiters in solar-system analogues, which would help in understanding exoplanet orbit alignments.
\end{abstract}

\keywords{photometry, robotic telescopes, surveys, large field-of-view}

\section{INTRODUCTION AND MOTIVATION}
\label{sec:intro}  

   \begin{figure} [ht]
   \begin{center}
   \includegraphics[width=0.9\linewidth]{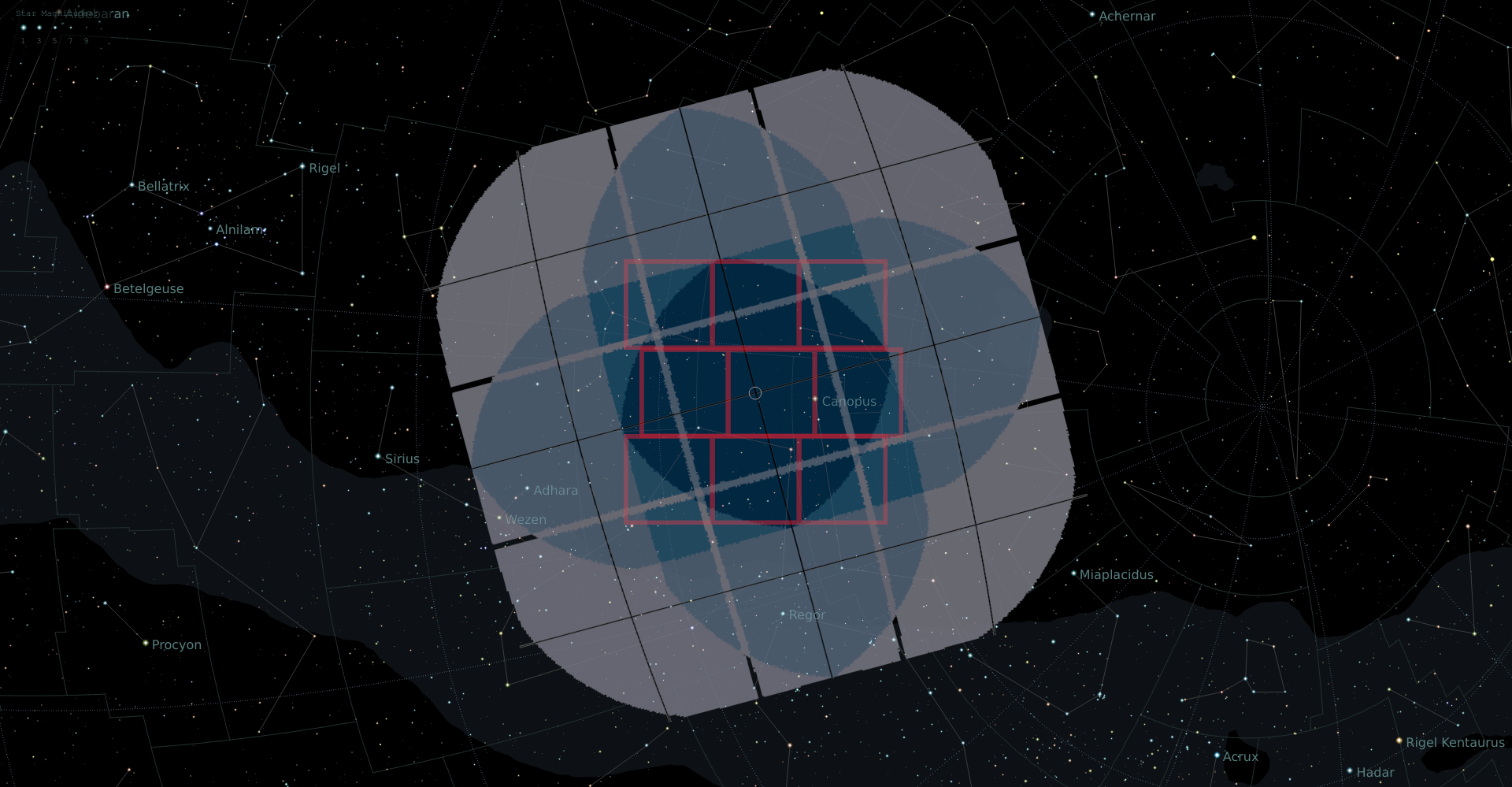}
   \end{center}
   \caption[fov of plato] 
   { \label{fig:skychart} 
The PLATO Southern field will be centered at $\alpha$=06:21:14.5, $\delta$=-47:53:13 and aligned parallel to the galactic plane. The four groups of six telescopes each are slightly tilted with respect to each other to allow for the extremely wide field of view of 2149~$^{\circ}$$^2$ square degrees. Only the central portion, approximately 325~$^{\circ}$$^2$ (dark blue) is covered by all 24 telescopes simultaneously and will deliver highest photometric precision. The BMK10k 3$\times$3 pointings are indicated with red squares. They have a small overlap of 100\,px (roughly 4 arc-min), except at Canopus, which is left out to avoid saturation in the detector.}
   \end{figure} 

ESA's M3 mission PLATO~\cite{2014ExA....38..249R} (PLAnetary Transits and Oscillation of stars) is the next-gen ultra-high-precision transit search for extrasolar planets. Its launch date is now confirmed for January 2027. Targeting much brighter stars in a much bigger field-of-view (FoV) than the extremely successful Kepler~\cite{2010Sci...327..977B} mission, it will ease ground-based spectroscopic follow-up and thus will provide not only radii, but also masses and hence average densities of the planets detected. It will also play a vital role in defining the most promising targets for planetary atmosphere characterization via transit spectroscopy on the ESO ELT. Eventually, it may help answering the question whether our solar system is special - or whether it is just a selection effect that no solar-system twin was detected to date.

PLATO's 24+2 telescopes -- where two telescopes are committed to fast read-out for astroseismology -- are grouped in six packs with their alignment slightly tilted to allow for an overlapping FoV. Two long-duration pointing fields were identified \cite{2025A&A...694A.313N}, with the ``Southern field'' located at $\alpha$=06:21:14.5, $\delta$=-47:53:13, as shown in Fig.~\ref{fig:skychart}. The main target class are dwarfs and subgiants with spectral types F5-K7. In the primary target sample, i.e., at a brightness limit V$\leq$ 11$^{\rm m}$ and a noise level below 50ppm, roughly 9,000 targets have been identified. The secondary target sample, containing identical stellar types as the primary sample, is for V$\leq$13$^{\rm m}$ but with more than 150,000 targets.

\begin{figure} [ht]
   \begin{center}
   \begin{tabular}{c} 
    \includegraphics[width=0.45\linewidth]{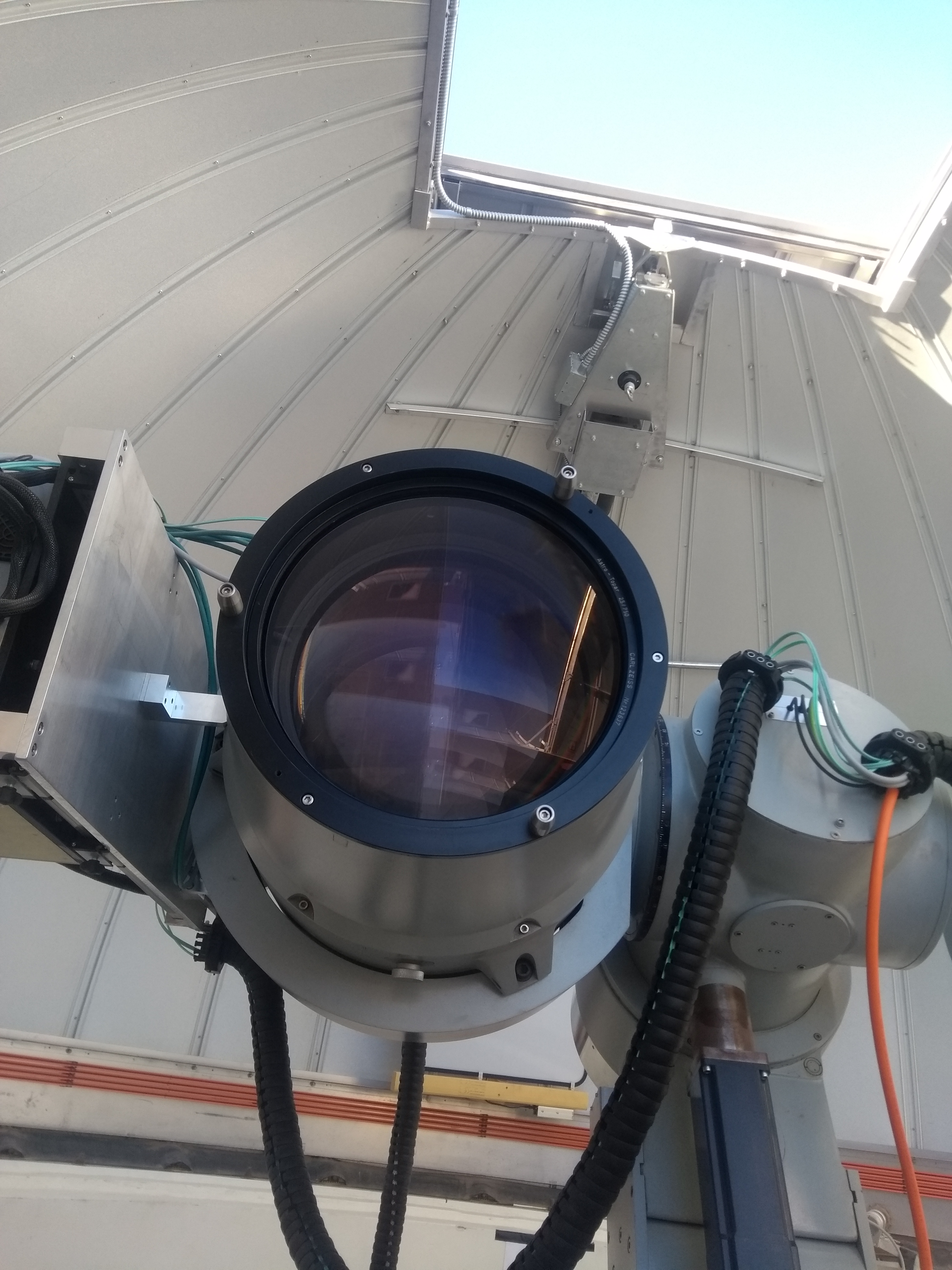}
    \includegraphics[width=0.45\linewidth]{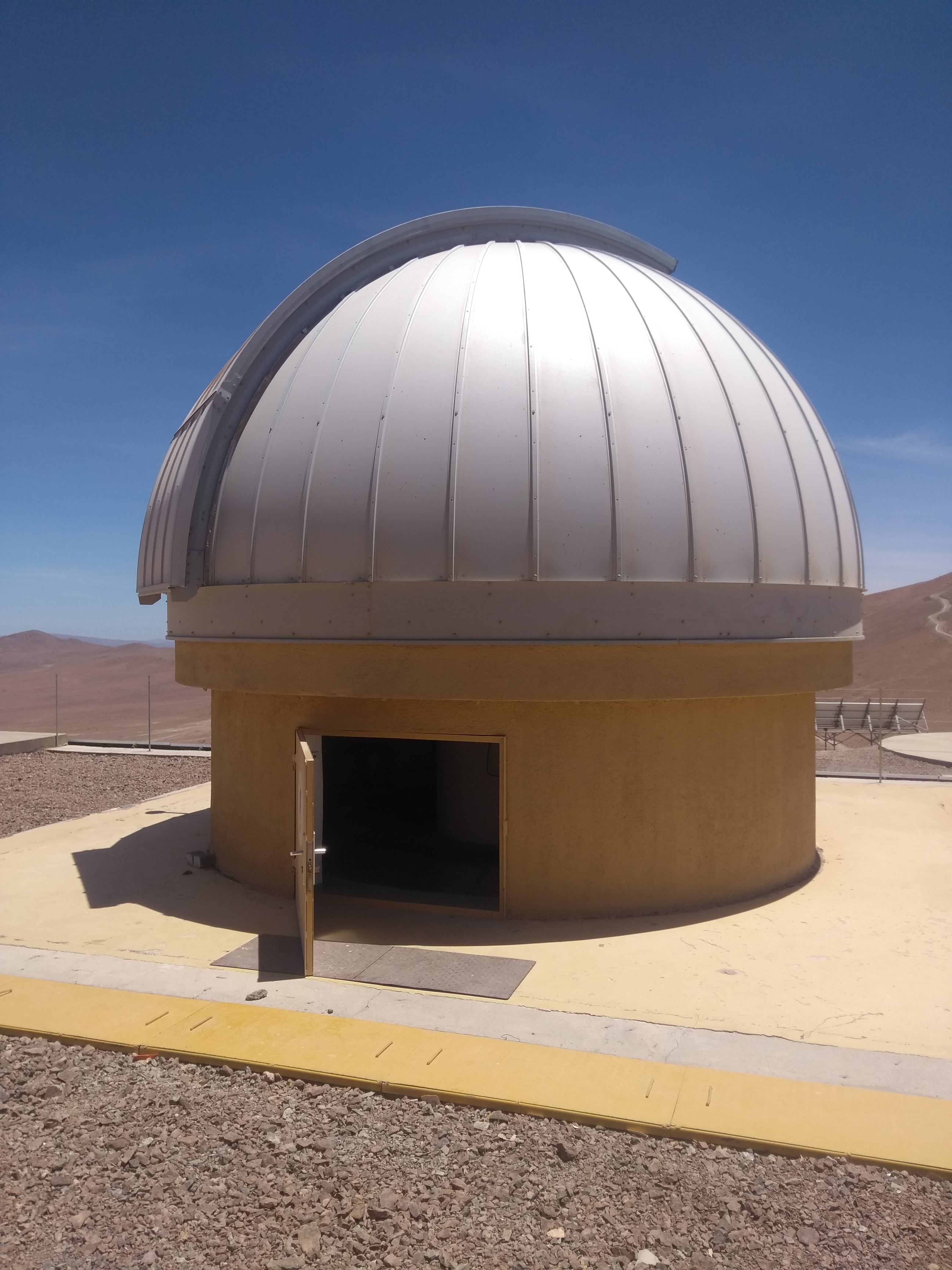}
   \end{tabular}
   \end{center}    \centering
    \caption{\label{fig:site} The BMK10k telescope on its new site, Cerro Murphy, which is a side-peak to Cerro Armazones.}
\end{figure}

The BMK10k's telescope, simply named BMK after Ballistische Messkammer\cite{1971MitAG..30..121S}, was built by Carl Zeiss Oberkochen in the 1970s and followed a design originally intended for the detection of ballistic missiles. Only two units were built, where the one now operated by AIP in Chile belongs to the Technical University of Munich and was used to photograph laser-illuminated satellites until 1990. It is now on loan to AIP and had been completely refurbished. The electronics were replaced with modern Beckhoff PLCs, the photographic plate exchanger was replaced with a larger CCD detector system, which was originally purchased as a spare for the PEPSI spectrograph's red arm~\cite{2015AN....336..324S}. The CCD, a STA1600LN device, is monolithic and was built by STA \cite{2012SPIE.8453E..1MB}. It has 10560x10560, 9$\mu$m pixel, providing a BMK10k FoV of 7.25$\times$7.25$^{\circ}$ at a sampling of 2.47"/pixel. It is read out in 50\,sec via 16 amplifiers, organized in a 8$\times$2 grid. The read-noise varies between 15 and 25 e$^{-}$; dark current is below two e$^{-}$ per pixel per hour at the nominal operating temperature of -90$^{\circ}$C. The cryogenic dewar is an identical copy to the original PEPSI design~\cite{2015AN....336..324S}. Peak quantum efficiency (QE) of 94\% is reached at 670\,nm and remains
above 90\% for 600–750\,nm but drops to 85\% at 550\,nm at
the central bandpass of Johnson~V, the fixed BMK10k filter realized at its CCD dewar-entrance window. The optical system of the telescope is an Astro-Topar lens system optimized for extremely high astrometric and photometric precision. The focal plane is planar over a linear range of 180$\times$180\,mm, almost twice the size of the detector; image distortion across the image diagonal is 5~$\mu$m \cite{1989sgmu.book.....S}. No focusing is possible, during commissioning the focus was measured and fixed with shimming plates to allow for a FWHM of roughly 2.5~px. The mechanical shutter consists of four rotating lamellas and allows exposure times down to 1/60~sec. The shortest flat exposures are limited to 1~sec., regular observations will probably never be shorter than 10~sec, thus, the expected shutter effects are small. For further details on the refurbishing and testing of the telescope we refer to~\cite{2019AN....340..712S}.

\section{SITE AND OBSERVING PLAN}
\label{sec:site}

\begin{figure}[ht]
   \begin{center}
   \begin{tabular}{c} 
    \includegraphics[width=0.45\linewidth]{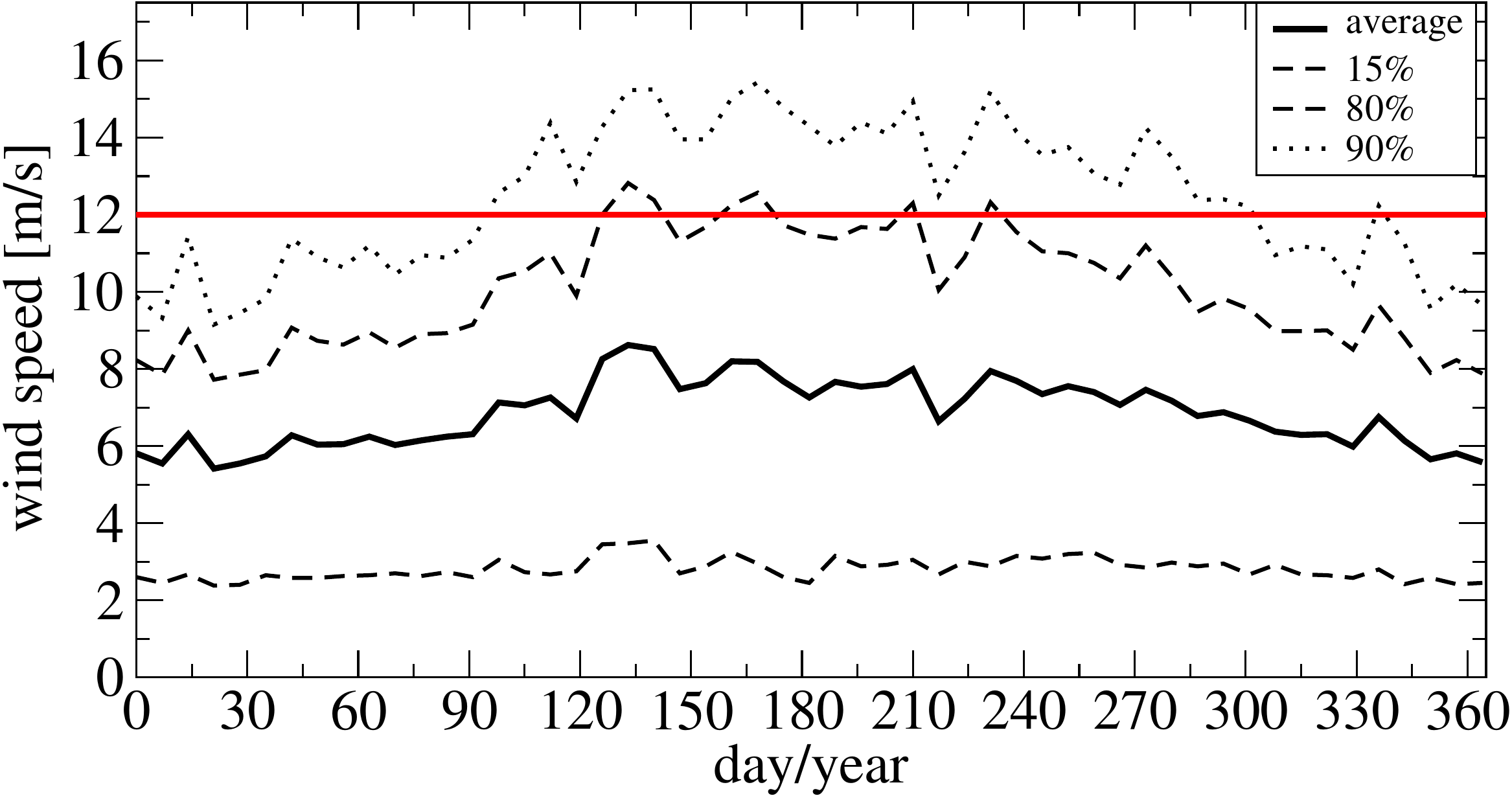}
    \includegraphics[width=0.45\linewidth]{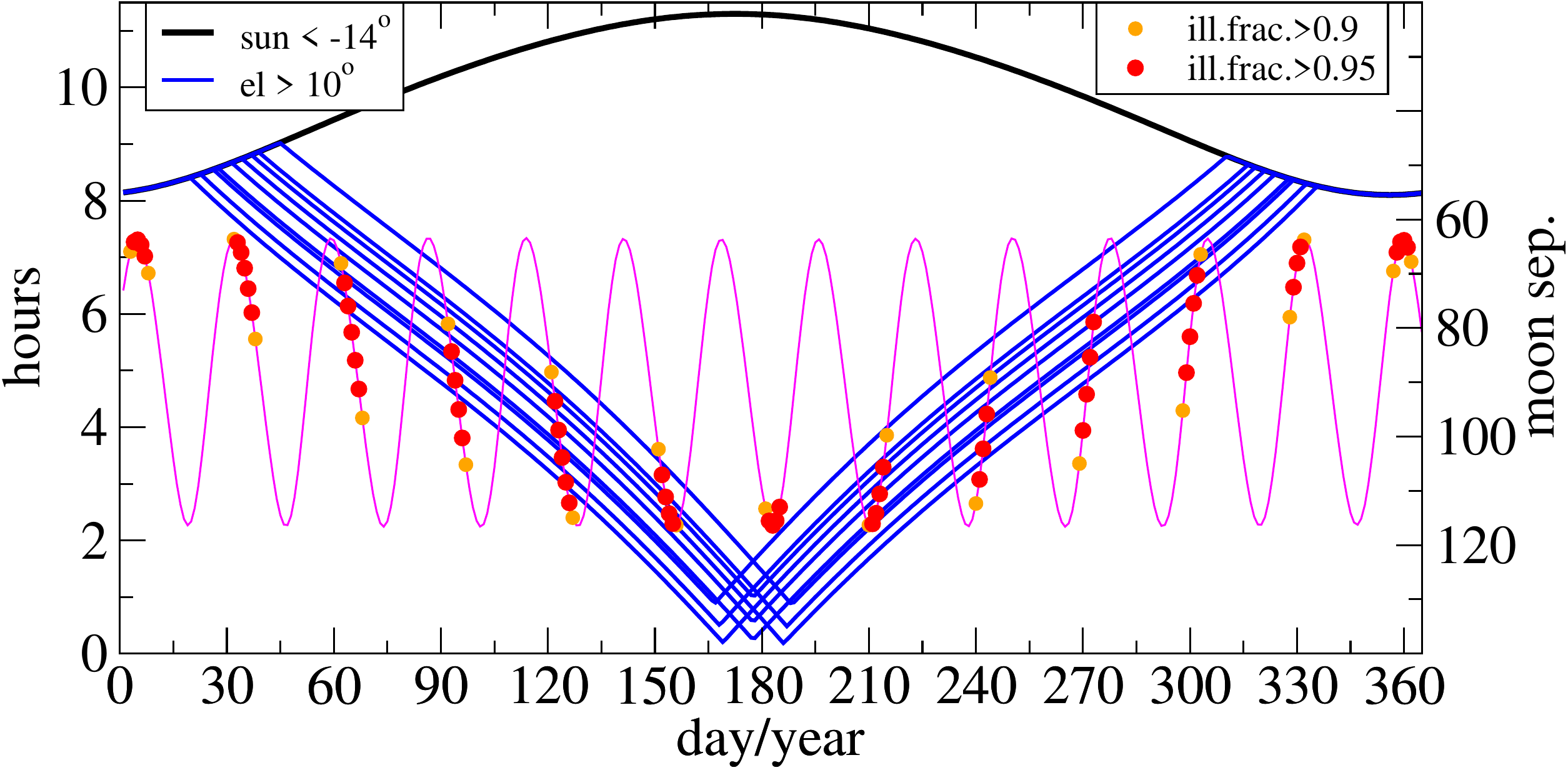}
   \end{tabular}
   \end{center}    \centering
    \caption{\label{fig:observe} To the left: seasonal wind speed statistics at Paranal from 1998-2026 plus 15\%, 80\%, and 90\% percentiles, respectively. The straight red line indicates our critical wind speed for observing stop. To the right: theoretical visibility limits of the 3x3 BMK10k PLATO fields. Even at solar conjunction the field remains visible for at least one integration.}
\end{figure}

The BMK10k is located at the National Polish Observatory at 70.20128$^{\circ}$W,  24.59867$^{\circ}$S and at an elevation of 2810\,m at a peak named Cerro Murphy (CMO=Cerro Murphy Observatory). This had been the location of the former Ruhr-University-Bochum (RUB) Observatory and is now part of ESO's ELT-site at Cerro Armazones. The new building with dome, see Fig.~\ref{fig:site}, was already completed in fall 2018, the BMK10k arrived on site in August 2019. The dome is a 5\,m ASH dome model REB; dome automation was implemented at AIP with a Beckhoff-PLCs already during initial commissioning in 2019. The entire project is laid out as a robotic facility, meaning no personnel is required on site. The entire operation of the site is managed by the Site Control System (SCS~\cite{2004AN....325..513G}), originally implemented for STELLA~\cite{2004AN....325..527S} on Tenerife. This software package is a generic package that allows operating any robotic facility; modifications are needed only at the ``driver'' base, i.e., individual parts of software that make the concrete hardware accessible to the generic workflow structure supported by SCS. The scheduling in the BMK10k case is rather simple: ensure that all 3$\times$3 fields are observed at tightest cadence possible, ensuring that the overall number of observations is the same for all fields. The location of the PLATO Southern field allows a continuous monitoring of all nine fields (see Fig.~\ref{fig:observe}, right panel). Even at times of conjunction to the Sun at least one visit is possible, owing to the Southern winter during June/July. The left hand side of Fig.~\ref{fig:observe} shows seasonal wind-speed averages from the Paranal site (retrieved from \verb|https://archive.eso.org/cms/eso-data/ambient-conditions/paranal-ambient-query-forms.html|, averaging done by the authors). It suggests that during Southern summer season, not more than 10\% of the time should be lost to high-wind circumstances. On the other hand, during winter season, where visibility of the PLATO Southern field is already sparse, up to 20\% of the observing time could be lost to high winds (precipitation and high humidity play a minor role). However, we want to note that wind-speeds are on average slightly higher at Cerro Murphy than on Paranal.

After a forced shut-down during the Covid pandemic, the site was electrically refurbished by the Nicolaus Copernicus Astronomical Center as part of the Polish Academy of Sciences in Warsaw. New solar panels were  installed and provide now enough power to operate the entire site without a connection to ESO's power grid. In April 2024 the entire site was officially re-opened. Second commissioning of the BMK10k took place in November 2024, resulting in first robotic light on Dec.\,9, 2024. Power hick-ups and difficult wind patterns led to a first commissioning data set that shows large time-gaps, but nevertheless allows for a first data quality assessment. In Souther spring (around October 2025), the rising temperatures revealed a problem with overheating of the azimuth motor. Observations had to be stopped until April 2026, when a new azimuth drive has been installed. The remainder of this paper covers lessons learned during commissioning.

\section{DATA REDUCTION}
\label{sec:data}

ESO guarantees a certain limit of data transfer per telescope, which is always sufficient to copy all science data fro the OCM to our home institute. However, currently it is also possible to transfer all calibration data, i.e., bias frames and twilight flat fields, to the home institution. The dome flat screen installed turned out to be unusable - at least at the current power supply situation - as the AC frequency of 50\,Hz introduces a changing pattern of bright lines in the flat screen. The number of twilight frames that can be taken were improved to 11-13 sky flats per twilight instead of the originally anticipated three~\cite{2019AN....340..712S}, but only at the cost of varying exposure times. For twilight flats, the telescope is pointed to the antisolar azimuth direction at a zenith distance of 25$^{\circ}$, which is the location with minimal sky-brightness gradient~\cite{1996PASP..108..944C}. Tracking is enabled, and the telescope is shifted slightly south by 6' after exposures. At dusk, sky flats start at a solar elevation of --8$^{\circ}$. At first, test images on a sub-window with the minimum exposure time of 1~sec are executed until the exposure level drops below an acceptable value (20,000~ADUs in our case). Once this is reached, full-frame readout is re-enabled; average count rates reached in the last flat-field exposure are used to anticipate the exposure time for the next flat field. During dusk, this means enlarging the exposure times until the maximum allowed flat-field exposure time of 10~sec is reached. Depending on the length of the twilight (shorter at equinox, longer at solstice), we can reach between 11 and 13 flat exposures per twilight with this procedure. Well before (respectively after) the twilight flats, a block of 25 bias frames is observed. Dark frames are currently omitted, as the dark signal at the nominal operating temperature is 1/30~e$^-$ at our current maximum science exposure time of 60~sec.

\subsection{Bias Treatment}

All bias frames taken undergo a sanity check on their average count rate and variance. This is also used as a double-check for CCD operating temperature and controller stability (increased temperature raises the average count rate, salt-and-pepper noise increases the variance). All 16 amplifiers come with an overscan section; this section is modeled with a fourth-order Lagrange polynomial and subtracted from the individual amplifier sections. The model shows RMS values below 1~ADU for all amplifiers. The overscan-corrected bias frames are then averaged, with outlier rejection to account for stochastic cosmics. This average master bias is then stored; the individual bias frames could be discarded at that point, but current band-width limits allow for a transfer of all bias frames back to the institute.

\subsection{Flat Treatment}

\begin{figure}[ht]
   \begin{center}
   \begin{tabular}{c} 
    \includegraphics[width=0.45\linewidth]{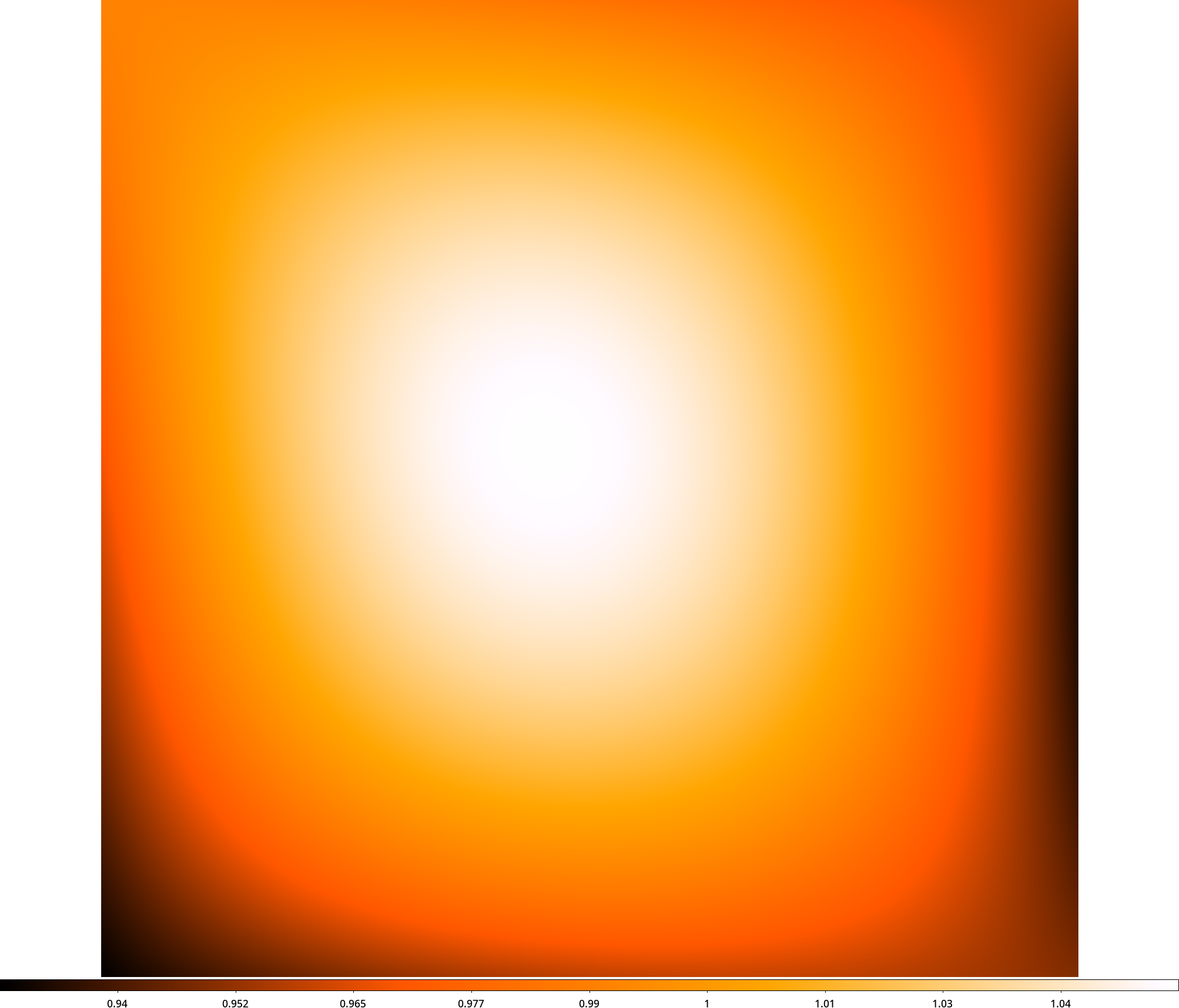}
    \includegraphics[width=0.45\linewidth]{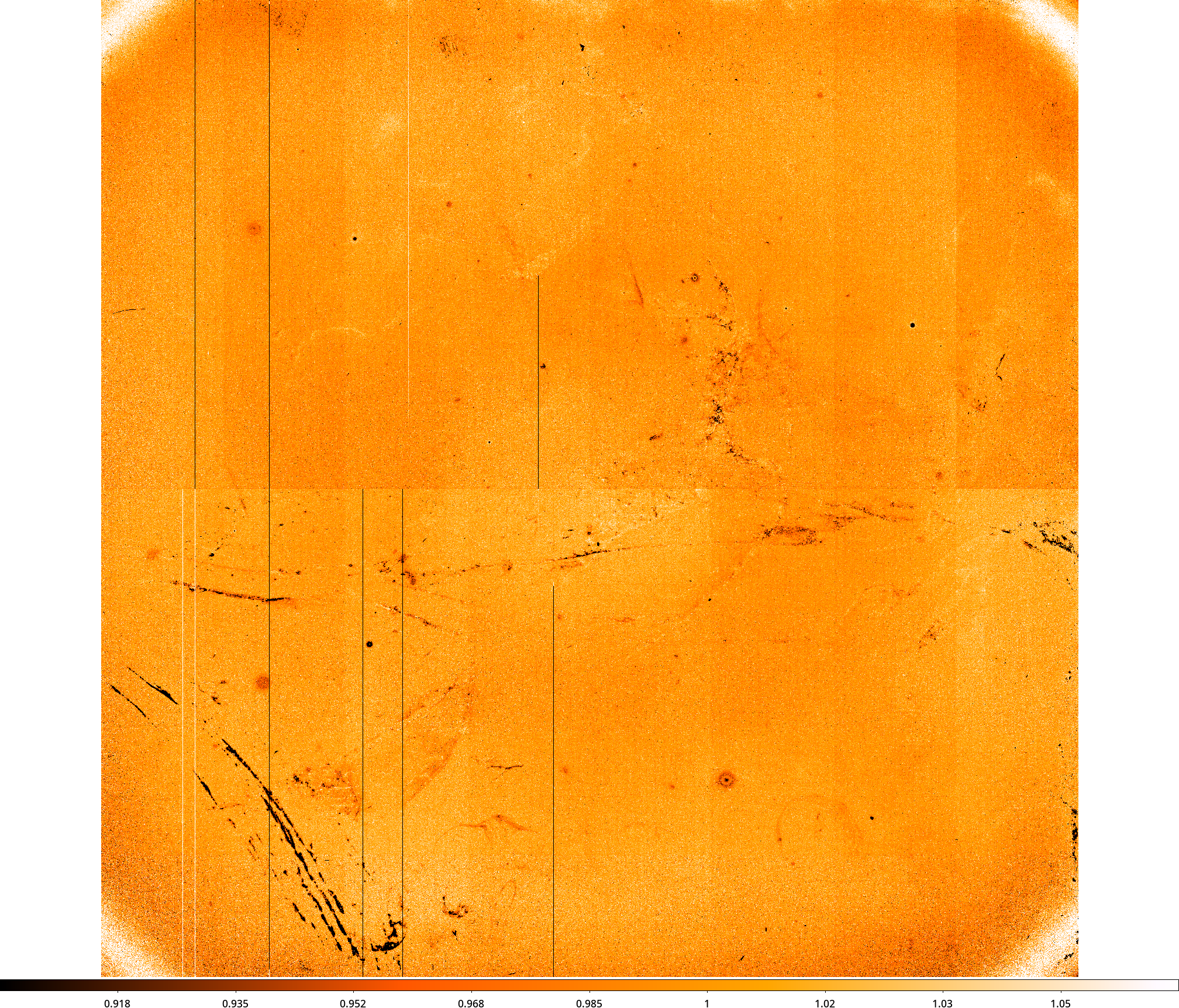}
   \end{tabular}
   \end{center}    \centering
    \caption{\label{fig:flat} To the left: a 2D Lagrange model to the stacked average flat field. To the right:  the residuals to the model which were used for final flat-fielding.}
\end{figure}

All flats gathered in a single twilight phase are combined in a two-staged process. Firstly, to account for different exposure levels, we define a global scaling factor $k$ such that $\chi^2$ is minimized. Twilight blocks with a $\chi^2$ exceeding a certain threshold are flagged as being affected by clouds and discarded.

\begin{equation}
\label{eq:optscale}
\chi^2=\sum_i^N(f_i-k\cdot g_i)^2 \, \rightarrow \, k=\frac{\sum_i^N f_i \cdot g_i}{\sum_i^N g_i^2},
\end{equation}

Secondly, cosmics and residual stellar signal are then filtered out at the second stage. For each pixel ADU$_{i,j}$ the global $k$'s are used to calculate an ADU sum and rejecting outliers by sigma-clipping. The amplifiers are gain-adjusted and the resulting flat is called the master flat.

For any given night, the closest ten successful master flats are first median combined, then $k$'s are calculated with respect to the median flat. If individual master flats exceed by a $\chi^2$ criterion, they are filtered out here. The final flat is called the average flat field of the night.

Large FoV twilight flats will always show a (varying) illumination gradient, even if pointed at the sky position of vanishing illumination gradient~\cite{1996PASP..108..944C}. Additionally, they tend to show light concentration in the CCD center, which can easily exceed the 10\% range. In such cases, care has to be taken not to decrease the central apparent efficiency. Wei \cite{2014SPIE.9149E..2HW} proposed to model the individual flats to a plane to compensate for inevitable sky-flat gradients. Dithered images to perform a proper illumination correction has been introduced by~\cite{1995A&AS..113..587M} and are used in many wide-field imagers (see, e.g., \cite{2015A&A...582A..62D}). We wanted to test another method: In the absence of vignetting, on can try to model the illumination with a 2D polynomial and take only the residuals as the pixel-to-pixel variations. This is done for the average flat field. The actual flat fields used are then the residuals to this model, normalized to the value of the coefficient to the zero-order Lagrange polynomial (i.e. the constant value). Fig~\ref{fig:flat} shows a third-order 2D Lagrangian model in the left panel and the residuals to the average flat in the right panel (note that for illustration purposes, the model was normalized). The center-to-limb variation is in the order of $1.04/0.94$ or 10\%, which translates roughly to a magnitude gradient of 0.1$^{\rm m}$. The residual, i.e., the flat field used to compensate pixel-to-pixel variations, appears flat, except for a residual systematic signal in the corners. The amplifier structure is still faintly visible, but always below the 1\% level. We think that the individual amplifier gains show small, long-term variations. We are currently investigating methods of mitigation. 

\subsection{Science frame treatment and WCS}

\begin{figure}[ht]
   \begin{center}
    \includegraphics[width=0.85\linewidth]{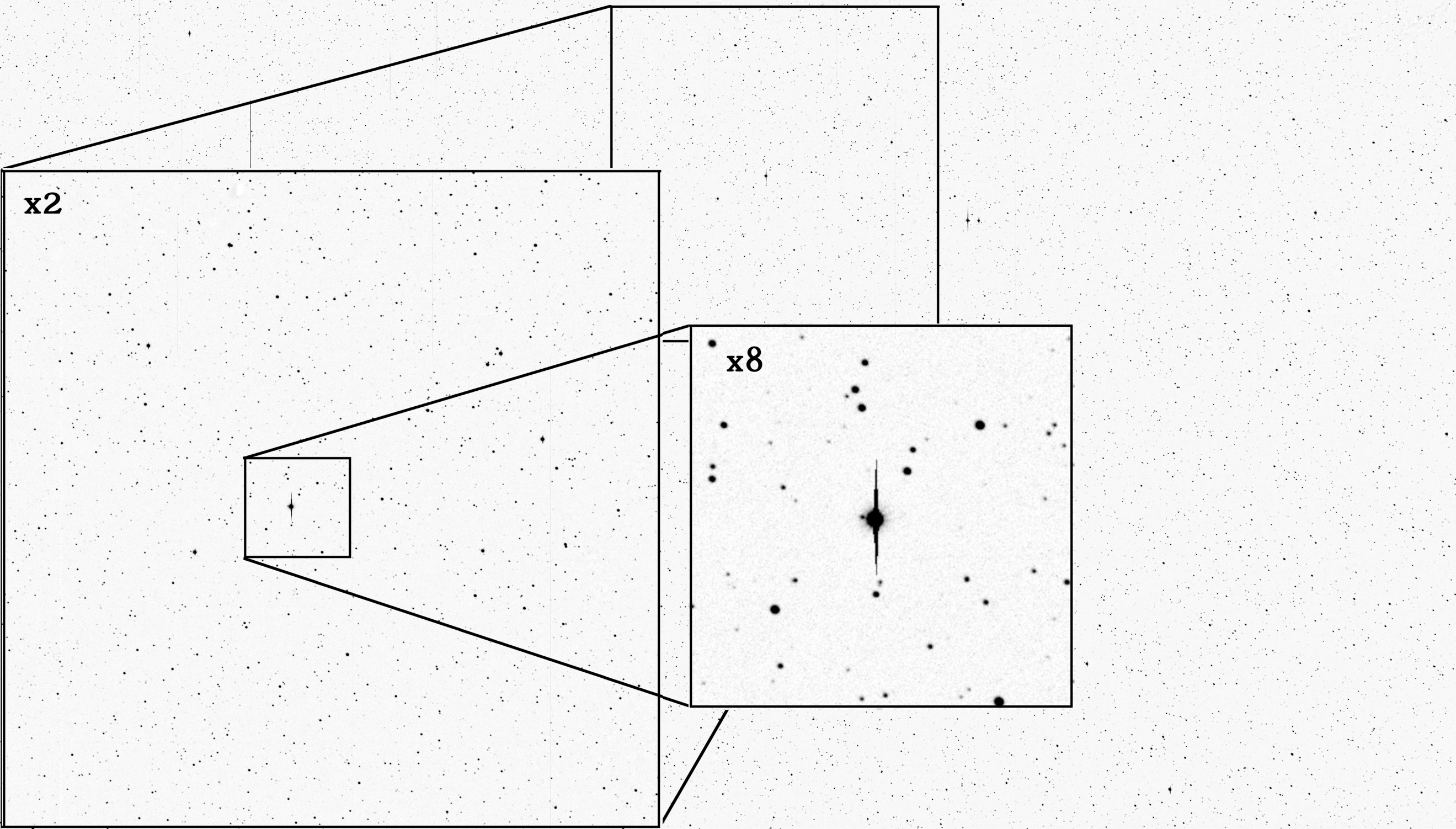}
   \end{center}    \centering
    \caption{\label{fig:zoom} Half of a single BMK10k CCD frame of the nine PLATO-South exposures. Exposure time was 60\,sec. The inserts are zooms by a factor 2 and 8, respectively. }
\end{figure}

Science frames are overscan-corrected. The bias is subtracted as the linear interpolation in time of the two master bias frames that bracket the night in question. Flat-fielding is done following the procedure outlined above, then the gain-corrected reduced amplifier sections are stitched together. Source-extractor~\cite{1996A&AS..117..393B} is used to extract the brightest stars, the World Coordinate System (WCS) is then solved using a local version of astrometry.net~\cite{2010AJ....139.1782L}. The stars found by source-extractor are matched to the GAIA DR3~\cite{2023A&A...674A...1G} catalog and the WCS is refined to the now $\approx$10,000 matched stars extracted. A simple zero-point determination together with a FWHM statistics is executed to detect obvious outliers. The science frames are put into the archive and are immediately available to enlisted users. It is still under discussion how public data releases are handled, the current plan is to make them public after one year of proprietary period.

\section{RESULTS FROM FIRST COMMISSIONING RUN}
\label{sec:results}

\begin{figure}[ht]
   \begin{center}
   \begin{tabular}{c} 
    \includegraphics[width=0.45\linewidth]{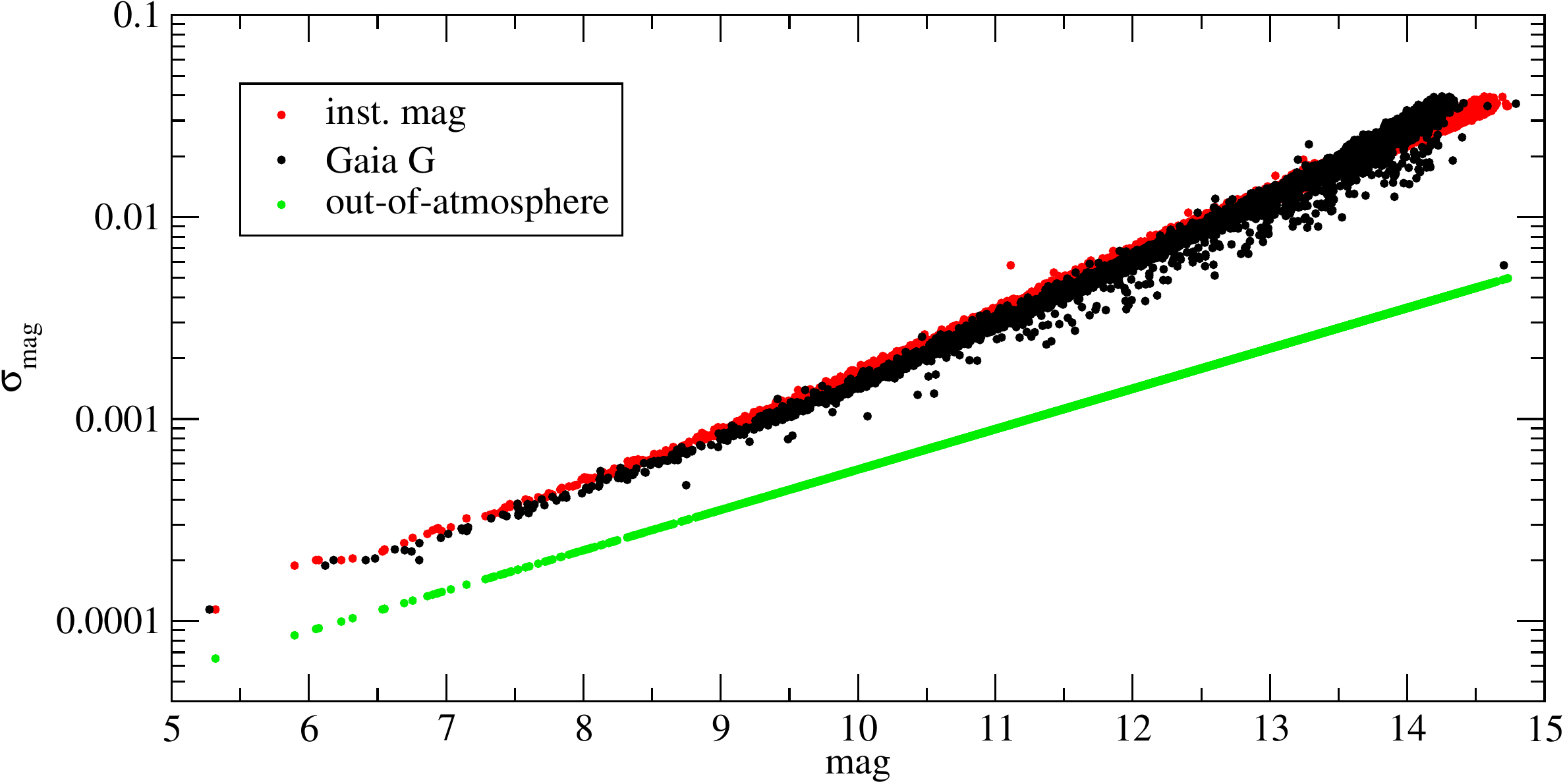}
    \includegraphics[width=0.45\linewidth]{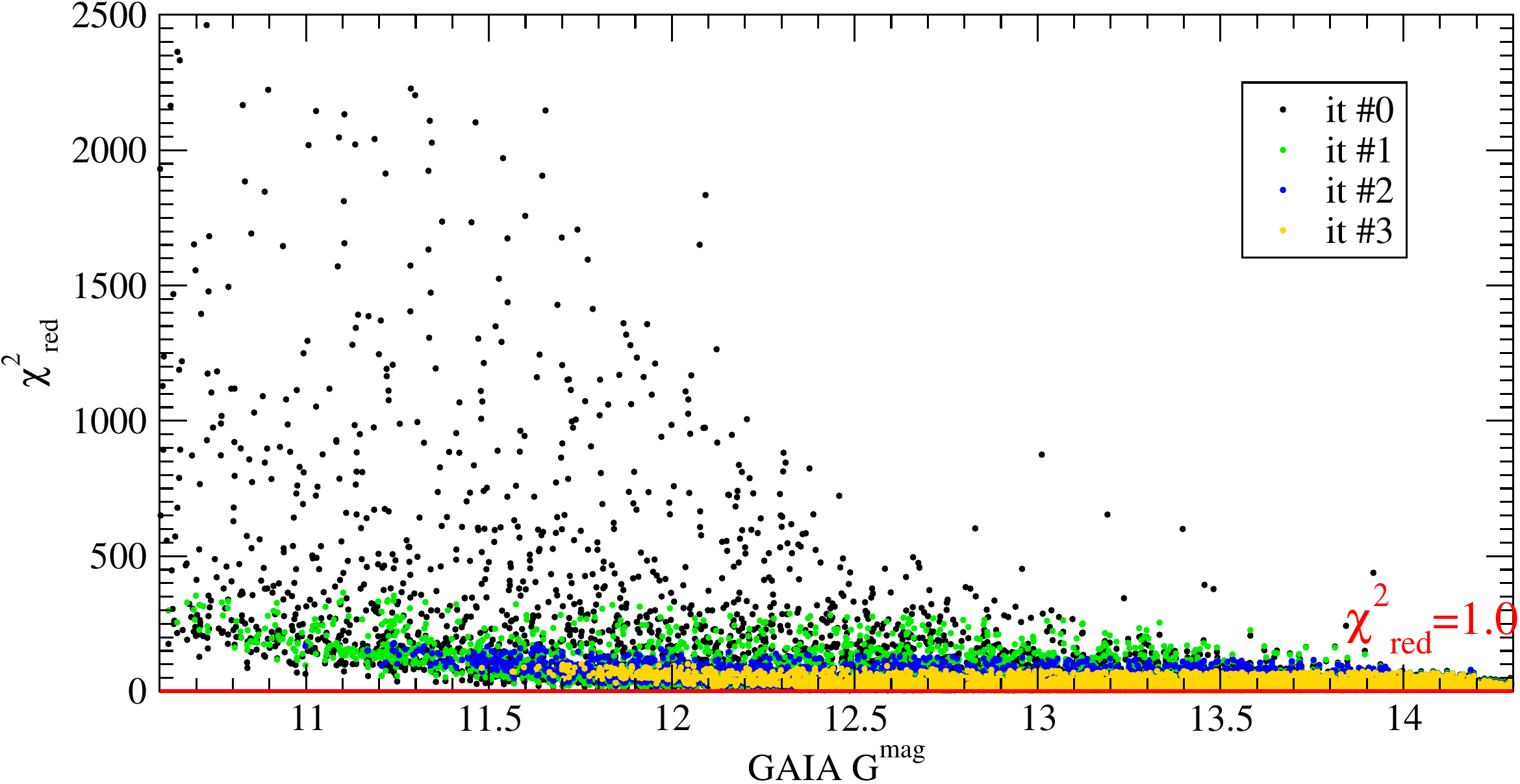}
   \end{tabular}
   \end{center}    \centering
    \caption{\label{fig:chi} To the left: a comparison of measured magnitude errors to theoretical, out-of-atmosphere, Poisson noise errors (green). The x-values of the black dots are the GAIA DR3 G$^{\rm mag}$ of the matched stars, while the red dots are instrumental magnitudes corrected for the zero-point. The right panel shows the reduced $\chi^2$ at the start of the process and after each iteration as a function of GAIA DR3 magnitudes of the matched stars.}
\end{figure}

As the number of detectable sources is close to 1 Mio. objects per frame, we devised a scheme for automated light-curve generation. We start out with creating an artificial comparison star by summing the flux of all stars which exceed a certain lower flux limit and also stay below an upper threshold, where we expect the detector to reach the non-linear regime. In an iterative process using reduced $\chi^2$ statistics (we settled on three times rejecting the top 10\% outliers), obviously variable stars are discarded, leaving a set of stars considered constant.

\begin{figure}[ht]
   \begin{center}
   \begin{tabular}{c} 
    \includegraphics[width=0.45\linewidth]{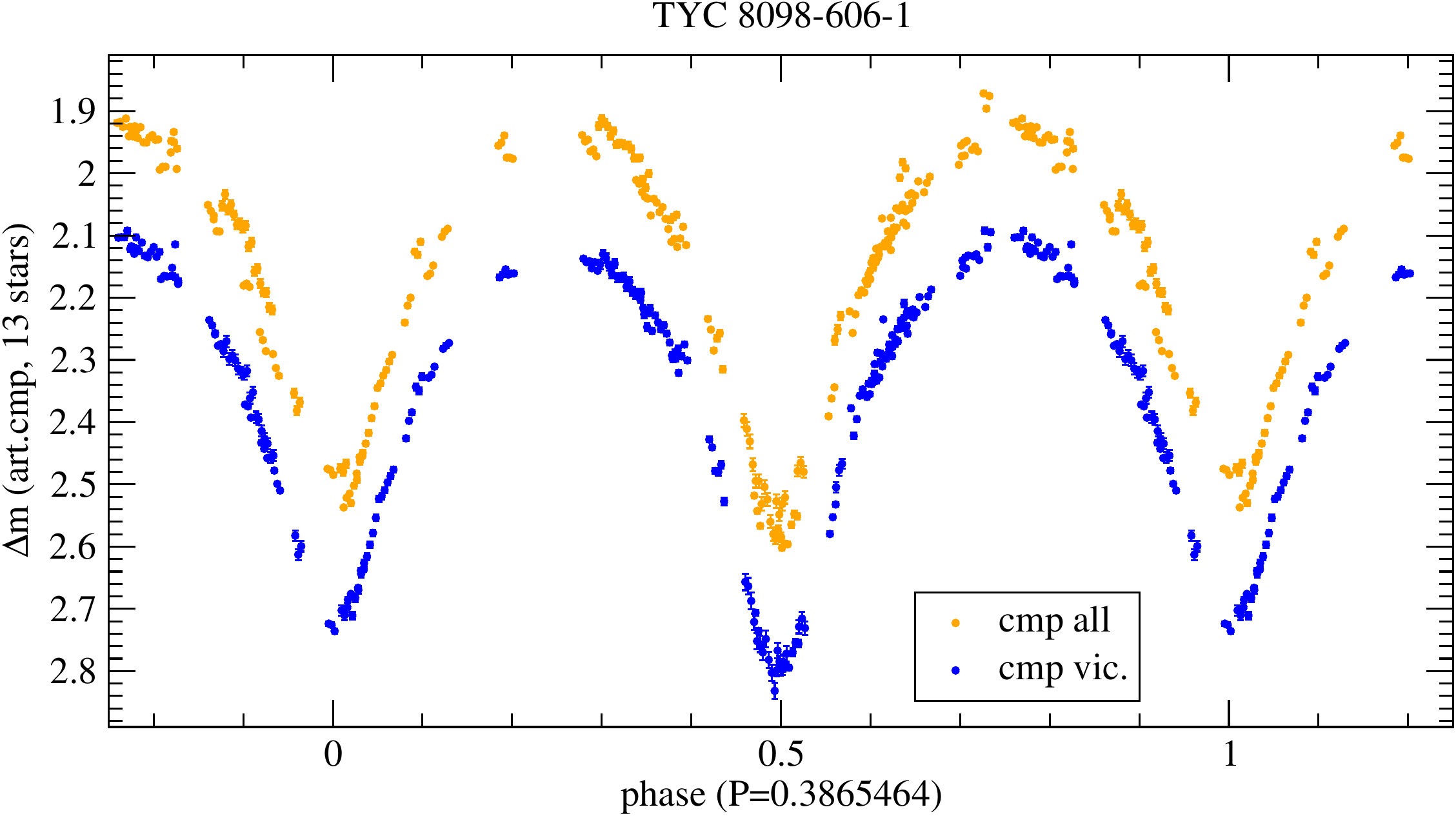}
    \includegraphics[width=0.45\linewidth]{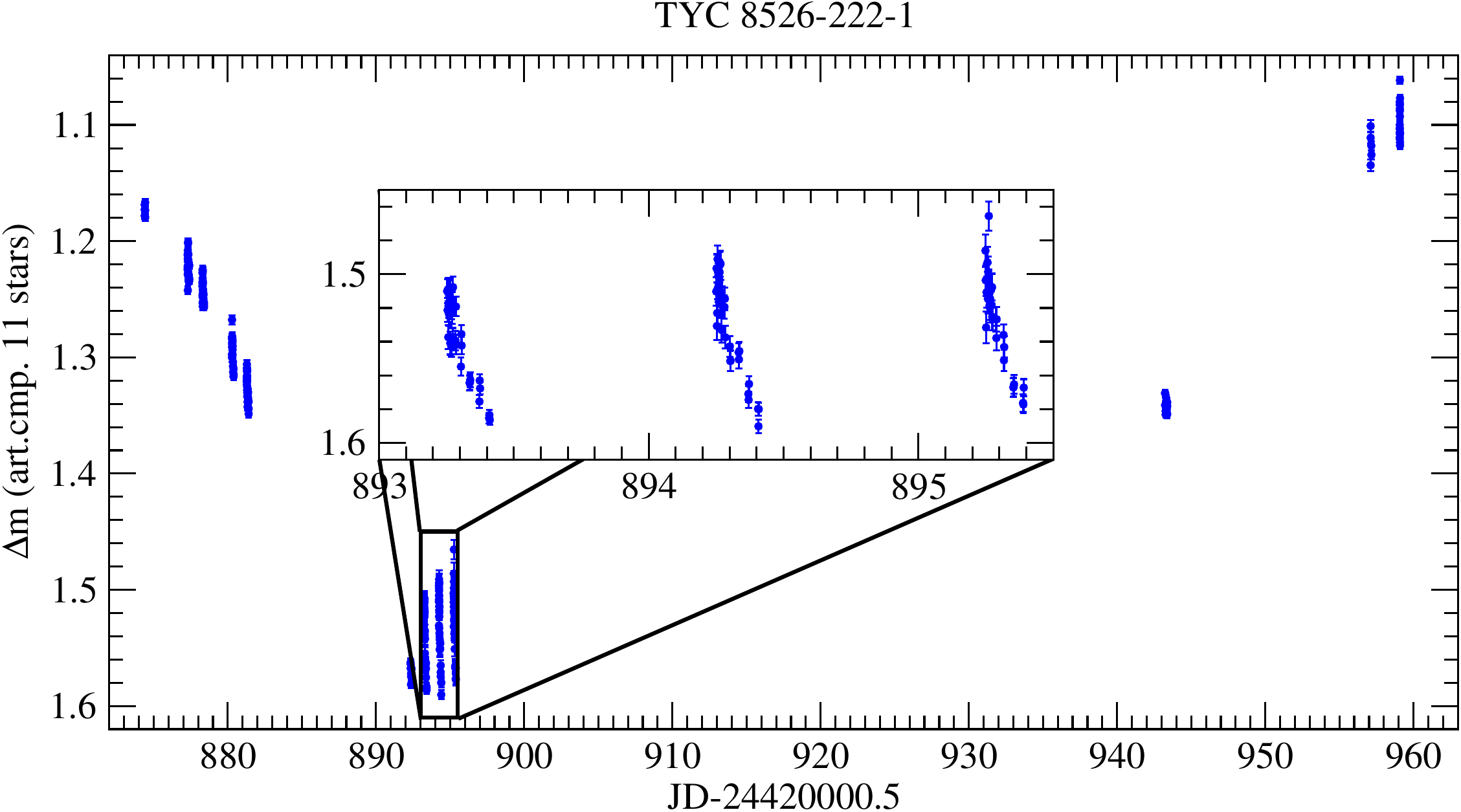}
   \end{tabular}
   \end{center}    \centering
    \caption{\label{fig:lc} To the left: TYC~8098-606-1, a known eclipsing binary at a period of P=0.3865464d. Blue points are magnitudes relative to a comparison constructed out of constant stars in the vicinity, in orange magnitudes are relative to all constant stars. To the right: TYC 8526-222-1, a suspected long-period variable, is a red star with BP-RP=2.60. It shows clear indication of remaining DR issues with extinction correction.}
\end{figure}

This set of constant stars can be used to readily construct differential light curves for the stars rejected as non-constants. However, due to the large FoV, extinction as a function of airmass $X$ and colour $(B-V)$ cannot be neglected: 

\begin{equation}
\label{eq:extinction}
m=m_{\rm inst}+\zeta+k'\cdot X+k"\cdot X \cdot (B-V)+...\, \rightarrow 
\Delta m=\Delta m_{\rm inst}+k'\cdot \Delta X +k"\cdot [X_i\cdot (B-V)_i -X_j\cdot (B-V)_j]+....
\end{equation}

It follows from Eq.~\ref{eq:extinction} using $\delta (k'\cdot X) \approx k' \cdot X \cot({h}) \cdot \delta h$  with $h$ the elevation of the object, that in the absence of a good knowledge of $k'$, better results are to be obtained at smaller $\delta h$. Currently, light curves are produced with a fixed $k'$=0.11 and $k"$=0, though it is well-known that both ``constants'' are dependent on atmospheric conditions and are known to vary in the order of 10\% over time. Thus, we expect better results if we limit the artificial stars to the summed flux of stars in vicinity of the variable in question. A maximum pixel distance of 250~px or roughly 10' was sufficient to always have at least a single constant star close to the variable.

In Fig.~\ref{fig:lc} two example light curves are shown. The left panel shows the phase folded light curve of TYC~8098-606-1, a known eclipsing binary at a period of P=0.3865464d. This plot illustrates the importance of using only local comparison stars: The blue curve is constructed for an artificial comparison consisting of 13 constant stars within 250~px of the target, while the orange curve shows differential magnitude with respect to an artificial comparison star comprised of all constant stars in the field. The second example, right-hand panel of Fig.~\ref{fig:lc}, shows the long-period variable candidate TYC 8526-222-1. This light curve exhibits a clear trend in its nightly variation with time, for the time of observing this directly translates to decreasing airmass, thus the star appears brighter at lower airmass even though it has been compared to stars in its vicinity only. Still, this is not really a surprise: the target star has a GAIA color of Bp-Rp=2.60 and is thus very red. Because longer wavelengths are less affected by extinction, a red star tends to appear \emph{brighter} than its surrounding (bluer) star at high air masses. As the pattern repeats itself from night to night, a tentative determination of $k"$ may be sufficient.

\section{CONCLUSIONS AND OUTLOOK}
\label{sec:conclusions}

The plot in Fig.~\ref{fig:chi}, right-hand panel, shows that the lowest reduced $\chi^2$ is achieved at magnitudes G$^{\rm mag}\ge 12$, with significantly worse 
$\chi^2$ at the magnitude range PLATO will achieve its best S/N. This may be due to our non-optimal aperture extraction for bright stars, but convinced us to revise the observing strategy to 2x10~sec plus 2x30~sec plus 2x60~sec for all nine BMK10k pointings. This will significantly extend the duration of a single observing block, but will give much better photometric precision at the bright edge. Currently, we rely on blind-pointing only. During commissioning, this led to varying pixel positions in the order of up to 200~px in either direction. Though polar alignment issues could be mainly fixed in the service mission in April 2026, we tend to include an acquisition phase at the start of each OB. Finally, extinction treatment: Assuming that $k'$ and $k"$ stay constant within a night, it should be possible to solve for these two external parameters considering \emph{all} constant stars in \emph{all} fields observed in a given night. This will be the route we will explore next.

\bibliography{export-bibtex} 
\bibliographystyle{spiebib} 

\end{document}